\documentclass{ws-ijmpd}
\usepackage[super,compress]{cite}
\usepackage{graphicx}
\usepackage{hyperref}

\begin{document}

\markboth{Debashree Sen}{Role of $\Delta$s in determining the properties of Neutron Stars in parameterized hydrostatic equilibrium}

%
\catchline{}{}{}{}{}
%

\title{Role of $\Delta$s in determining the properties of Neutron Stars in parameterized hydrostatic equilibrium}

\author{Debashree Sen\footnote{p2013414@goa.bits-pilani.ac.in}}

\address{Birla Institute of Technology and Science-Pilani, K K Birla Goa Campus, NH-17B, Zuarinagar, Goa-403726, India}

\maketitle

\begin{history}
\received{Day Month Year}
\revised{Day Month Year}

\end{history}

\begin{abstract}

The possible existence of $\Delta$ resonances is inspected in the cold dense matter of neutron star (NS) core in presence of the hyperons. The diverse effects of variation in $\Delta$ mass on their formation and the equation of state (EoS) are studied in this work with an effective chiral model and the resultant NS properties are calculated with the help of parameterized Tolman-Oppenheimer-Volkoff equations (PTOV) to bring out the two important features of pressure in the context of massive NSs. The $\Delta$ puzzle is re-explored and resolved taking into account the concept of modified/parameterized inertial pressure and self-gravity in case of massive pulsars like PSR J1614-2230 and PSR J0348-0432. It is seen that although the presence of exotic matter like the hyperons and the $\Delta$s softens the EoS considerably, their presence in massive NSs can be successfully explained with the theory of parameterized hydrostatic equilibrium conditions. The results of this work also satisfy the constraints on $R_{1.4}$ and $R_{1.6}$ from the gravitational wave (GW170817) detection of binary NS merger. The constraint on baryonic mass from PSR J0737-3039 is also satisfied with the solutions of the PTOV equations for all the $\Delta$ masses considered.   

\end{abstract}

\ccode{PACS numbers: 26, 21.65.+f, 12.39.Fe, 97.60.Jd, 26.60.+c, 26.60.−c, 14.20.Jn, 04.50.Kd}

\section{Introduction}
\label{intro}

 Born as remnants of core-collapse supernova, the neutron stars (NSs) provide the best conditions for the study of dense matter \cite{Glen,Glen5,Glen6,Glenq}. It is known that the density of the core of NSs ranges upto a few times normal nuclear matter density ($\rho_0 \sim 0.16~fm^{-3}$). However, despite of a huge amount of investigations, the presence of exotic matter such as hyperons, $\Delta$ isobars, quarks and types of bosonic condensates etc. at such conditions, still remains inconclusive. At present the insufficient understanding of nuclear interactions in this density domain leads to uncertainty in the equation of state (EoS) which largely determines the composition and structure of the NSs. However, the EoS is strongly constrained with the discovery of massive NSs like PSR J1614-2230 ($M = (1.928 \pm~ 0.017) M_{\odot}$) \cite{Fonseca} and PSR J0348+0432 ($M = (2.01 \pm 0.04) M_{\odot}$) \cite{Ant}. The recent detection of gravitational wave (GW170817) of binary neutron star (BNS) merger \cite{Abbott} suggest that the canonical radius ($R_{1.4}$) lies within the range $R_{1.4} < 13.76$ km \cite{Fattoyev} or $12.00 \leq R_{1.4} \leq 13.45$ km \cite{Most}. Also it has been suggested from the BNS coalescence that the upper limit on the radius of a 1.6 $M_{\odot}$ NS is $R_{1.6} \leq 13.3$~km \cite{Abbott,Fattoyev} while its lower limit is prescribed to be $R_{1.6} \geq 10.68_{-0.04}^{+0.15}$~km \cite{Bauswein,Fattoyev}. Satisfying these constraints in the presence of hyperons and $\Delta$ baryons therefore becomes challenging since it is well known that with these additional degrees of freedom the EoS softens considerably yielding low mass NSs. This gives rise the hyperon and the $\Delta$ puzzles.
 
 In this work I investigate the possibility of formation of $\Delta$ resonances ($\Delta^{-,0,+,++}$) in neutron star matter (NSM) in the presence of the hyperons ($\Lambda$ (1116 MeV), $\Sigma^{-,0,+}$ (1193 MeV) and $\Xi^{-,0}$ (1318 MeV)). It is well known that the $\Delta$s posses a Breit-Wigner mass distribution having the centroid mass at 1232 MeV and a width of about 120 MeV \cite{prc92,Sun}. Therefore in this work I look for the effect of variation in mass of the $\Delta$ baryons on the population of different particles in NSM, the EoS and the resulting global properties of NSs. A lot of work has been done to resolve the hyperon puzzle \cite{Miyatsu,Stone,Dhiman,Dexheimer,Bednarek,Bednarek2, Weissenborn12,Weissenborn14,Agrawal,Lopes, Oertel,Colucci,Dalen,Lim,Rabhi,TKJ2,TKJ3,Sen2,Sen3} and the $\Delta$ puzzle \cite{prc92,Sun,Kolomeitsev,Maslov,Zhu,Zhou,Torsten,del1,del4, DragoPRC,Chen,Lavagno,Hu,Oliveira,Oliveira2,Oliveira3, Rodrigues,JiaJieLi,Sahoo,Sen}. The formation of the $\Delta$s and the hyperons in NSM are strongly controlled by their respective couplings with the mesons which are at present indefinite for the $\Delta$s. In our previous work \cite{Sen} we have already discussed in details the sensitivity of formation of the $\Delta$s to the individual scalar, vector and the isovector mesons and their effects on the NS properties. In absence of any experimental data for the $\Delta$-meson couplings, we choose the scalar and vector meson couplings with the $\Delta$s in accordance with the theoretical prescription for the same from QCD calculations \cite{qcd}. It has already been shown that the formation of the $\Delta$s and the properties of the NSs are most sensitive to the isovector coupling for the $\Delta$s \cite{prc92,Sen}. The increase in the isovector coupling reduces the concentration of the $\Delta$s in NSM and increases the mass of the NS \cite{prc92,Sen}. Hyperon couplings, on the other hand, are well established by the various coupling schemes like those based on the SU(6) \cite{Bednarek,Weissenborn12,Katayama,Schaffner-Bielich,Sulaksono,Bhowmick,Sahoo2} or SU(3) \cite{Miyatsu,Weissenborn11,Weissenborn13,Miyatsu2} quark models and constrained by certain hypernuclear studies \cite{Glen,Glen5,Glen6,TKJ2,TKJ3,Rufa,Arumugam} and the individual hyperon potential depths ($\Lambda$~(-28 MeV), $\Sigma^{-,0,+}$ (+30 MeV) and $\Xi^{-,0}$ (-18 MeV)) \cite{Schaffner-Bielich,Sulaksono,be}.

 I therefore fix the hyperons and $\Delta$ couplings consistent with the aforesaid constraints on the same and investigate the their formation in NSM with variation in $\Delta$ baryon mass $m_{\Delta}=1232\pm120$ MeV \cite{prc92,Sun} within the effective chiral model \cite{TKJ2,TKJ3,Sen2,Sen3,Sen,Sahu,TKJ,TKJ4}. The model is based on chiral symmetry and dynamical mass generation of the baryons and the scalar and vector mesons in terms of the vacuum expectation value (VEV) of the scalar field. The model along with the chosen parameter set is well tested and co-related with saturated nuclear matter properties \cite{TKJ2,TKJ3,Sen,Sen2,Sen3,TKJ,TKJ4}. It is therefore the model provides very less number of free parameters for the adjust of nuclear saturation properties. Having obtained the EoS with the hyperons and the $\Delta$s, the global properties of the NSs like the central density, gravitational mass, baryonic mass and the radius are calculated in static conditions with the parameterized hydrostatic equilibrium conditions as suggested by \cite{Velten}.
 
 In the present work I intend to investigate the possibility of obtaining massive NS configurations in the presence of hyperons and the $\Delta$s with varied $\Delta$ mass. Apart from the hadron-quark phase transition \cite{Ozel,Weissenborn11,Klahn,Bonanno,Lastowiecki,Dragoq1, Dragoq2,Bombaci16,Zdunik,Masuda16,Wu,Wu1} or the hyperon-hyperon interaction via exchange of vector mesons \cite{Bednarek,Weissenborn12,Oertel,Maslov} or scalar mesons \cite{Dalen}, there are many recent paradigm associated with modifications/extensions to the normal General Relativity (GR) theory to resolve the hyperon and the $\Delta$ puzzles. Such modified gravity theories conclude that massive pulsars like PSR J1614-2230 (M = (1.928 $\pm~ 0.017) M_{\odot}$) \cite{Fonseca} and PSR J0348+0432 (M = (2.01 $\pm~ 0.04) M_{\odot}$) \cite{Ant} are objects with massive gravity, for which ordinary General Relativity (GR) may not be a sufficient approach. Many of these theories also suggest that massive NSs can also constrain gravity along with the EoS \cite{Velten,Doneva,Eksi,Oliveira4,Brax}. Certain extended/modified theories of gravity like f(R) gravity \cite{Brax,Capozziello,Sotiriou,Arapoglu,Yazadjiev,Staykov,YazadjievPRD91,Santos,Astashenok}, scalar-tensor theories \cite{Brans,Damour,Minamitsuji,Freire,Ramazanoglu,Yazadjiev2,Cisterna,Maselli,Silva,Harada,Novak,Salgado, Sotani,Pani}, quadratic gravity like Einstein-dilaton-Gauss-Bonnet gravity \cite{Yagi,Kleihaus} and Chern-Simons gravity \cite{Alexander,Ali-Haimoud,Yunes,Yagi2}, extended theories of gravity \cite{Wojnar,Wojnar2,Burikham}, massive gravity \cite{Hendi} has been used to modify the general Tolman-Oppenheimer-Volkoff (TOV) equations \cite{tov,tov1} to calculate the properties of NSs. Some works \cite{Velten,Glampedakis,Glampedakis2,Schwab} have shown the success of parameterization of the TOV equations for the same purpose, without involving any particular theory of gravity. In order to calculate the gross NS properties, I adopt the same parameterized TOV (PTOV) equations as suggested by \cite{Velten}. In the present work, the parameterized hydrostatic equilibrium conditions are considered in terms of the two most important parameters in connection to pressure - the one ($\beta$) coupling with inertial pressure to contribute to the overall mass density and the other ($\chi$) affecting the gravitational effects of pressure, known as the self-gravity of the star \cite{Eksi,Schwab}. These parameters and their variations are physically justified and may play crucial role in supporting high mass NS configurations \cite{Sen3,Velten,Schwab}. Representing a potentially crucial modification to GR, the PTOV equations are universally applicable to all EoS specially to those obtained from certain RMF models that do not fulfill the 2 $M_{\odot}$ mass constraint of NSs despite satisfying all the saturated nuclear matter properties \cite{Dutra}. Such EoS can be subjected to the PTOV equations to achieve the massive NS configurations. This ensures that the self-gravity of the star is equally important as the EoS to contribute to its total gravitational mass. These PTOV equations are not based on any particular theory of modified/extended gravity but the parameters and their variation are not arbitrary and carry their individual physical significance. In a recent work, \cite{Sen3} we presented a possible bound on these two parameters within the framework of our effective chiral model including the presence of hyperons and showed that these modifications may depict how the effects of pressure deviates from normal GR conditions in case of massive NSs. Therefore, in the present work, I test the effectiveness of the PTOV equations to generate high mass NS configurations, consistent to observational constraints, with the addition of further exotic degrees of freedom, the $\Delta$s along with the hyperons.
 
  The present manuscript is organized in the following manner. In section 2 we describe our effective chiral model including the baryon decuplet ($\Lambda$, $\Sigma^{-,0,+}$, $\Xi^{-,0}$, $\Delta^{-,0,+,++}$) along with the nucleons (n,p). We also specify our model parameters and the coupling scheme adopted for the hyperons and $\Delta$ to the mesons. The basic formalism to obtain static properties of NSs using both general and PTOV equations are also discussed in this section. The results obtained are shown and discussed in section 3. The final conclusions of the work are mentioned in section 4.

\section{FORMALISM}
\subsection{Effective Chiral Model with baryon octet and the Delta baryons}

 The phenomenological model \cite{Sahu,TKJ,TKJ2,TKJ3,Sen,Sen2,Sen3} deals with the manifestation of chiral symmetry and its spontaneous breaking at ground state due to which the scalar field $\sigma$ attains a VEV $\sigma_0=x_0$. The scalar $\sigma$ and the pseudoscalar $\pi$ mesons are the chiral partners and $x^2 = ({\pi}^2+\sigma^2)$ \cite{VolKo}. The vector $\omega$ meson was introduced to the model in \cite{Sahu} with a dynamically generated mass in terms of $x_0$ due to interaction of the scalar ($\sigma$) and the pseudoscalar ($\pi$) mesons with the isoscalar vector boson ($\omega$). The effective Lagrangian density \cite{Sahu,TKJ,Sahu2,Sen,Sen2,Sen3} for the effective chiral model is given by
 
\begin{eqnarray}
\hspace*{-2.0cm}\mathcal{L} = \overline{\psi}_B \left[\left(i \gamma_{\mu} \partial^{\mu} - g_{\omega B}~ \gamma_{\mu} \omega^{\mu}  -\frac{1}{2} g_{\rho B}~ 
\overrightarrow{\rho_{\mu}} \cdot \overrightarrow{\tau_B} \gamma^{\mu} \right)-g_{\sigma B} 
\left(\sigma + i \gamma_5 \overrightarrow{\tau_B} \cdot \overrightarrow{\pi} \right) \right] \psi_B \nonumber \\ 
\hspace*{-1.0cm}+ \frac{1}{2} \left(\partial_{\mu} \overrightarrow{\pi} 
\cdot \partial^{\mu} \overrightarrow{\pi} + \partial_{\mu} \sigma ~ \partial^{\mu} \sigma \right)
-{\frac{\lambda}{4}} \left(x^2-x_0^2\right)^2 - \frac{\lambda B}{6} (x^2-x_0^2)^3 
- \frac{\lambda C}{8}(x^2-x_0^2)^4 \nonumber \\
-\frac{1}{4}F_{\mu\nu}F^{\mu\nu} +\frac{1}{2}\sum_B g_{\omega B}^2~x^2~\omega_\mu 
\omega^\mu -\frac{1}{4}~\overrightarrow{R_{\mu\nu}} \cdot \overrightarrow{R^{\mu\nu}}
+\frac{1}{2}~m_\rho^2 ~\overrightarrow{\rho_\mu} \cdot \overrightarrow{\rho^\mu} 
\protect\label{Lagrangian}
\end{eqnarray}

$\psi_B$ is the baryon spinor and the subscript B denotes sum over all baryonic states viz. the nucleons and the hyperons (sumover index $B = n, p, \Lambda, \Sigma^{-,0,+}, \Xi^{-,0}, \Delta^{-,0,+,++}$). B and C are the higher order scalar couplings. The nucleons (N=n,p), the hyperons (H=$\Lambda$,$\Sigma^{-,0,+}$, $\Xi^{-,0}$) and the deltas ($\Delta^{-,0,+,++}$) interact with eachother, mediated by the scalar $\sigma$ meson, the vector $\omega$ meson (783 MeV) and the isovector $\rho$ meson (770 MeV) with respective coupling strengths $g_{\sigma_B}, g_{\omega_B}, g_{\rho_B}$. These couplings along with the higher order scalar couplings B and C are evaluated at nuclear saturation density $\rho_0 = 0.153~fm^{-3}$. The masses of the baryons ($m_B$) and the scalar and vector mesons can be expressed in terms of $x_0$ \cite{Sahu,Sahu2,TKJ,Sen,Sen2,Sen3} as

\begin{eqnarray}
m_B = g_{\sigma_B} x_0,~~ m_{\sigma} = \sqrt{2\lambda} ~x_0,~~
m_{\omega} = g_{\omega_N} x_0
\end{eqnarray}

where, $\lambda=({m_\sigma}^2 -{m_\pi}^2)/(2{f_\pi}^2)$ is derived from chiral dynamics. $f_\pi$, being the pion decay constant, relates to the vacuum expectation value of $\sigma$ field as $<\sigma>=\sigma_0~=f_\pi$ \cite{TKJ}. Since in the relativistic mean field treatment (MFT), $< \pi >= 0$ and the pion mass is $m_{\pi}=0$, the explicit contributions of pions are neglected \cite{TKJ,TKJ2,TKJ3}.
 
 The isospin triplet $\rho$ meson accounts for the asymmetric nuclear matter. Its coupling strength is obtained by fixing the symmetry energy coefficient $J = 32$ MeV at $\rho_0$, given by

\begin{eqnarray}
J = \frac{C_{\rho_N}~ k_{FN}^3}{12\pi^2} + \frac{k_{FN}^2}{6\sqrt{(k_{FN}^2 + m_N^{\star 2})}}
\end{eqnarray}

 where, $C_{\rho_N} \equiv g^2_{\rho_N}/m^2_{\rho}$ and $k_{FN}=(6\pi^2 \rho_N/{\gamma})^{1/3}$.


 Using the relativistic MFT \cite{Glen,MulSer}, the equation of motion (at T=0) for the fields are calculated. The VEVs of $\omega$ and $\rho$ fields in terms of the total baryon density $\rho$ is given as
 
\begin{eqnarray}  
\omega_0 = \frac {\sum\limits_{B} g_{\omega_B} \rho_B}{\Bigl(\sum\limits_{B} g_{\omega_B}^2 \Bigr) x^2}
\label{vector_field}
\end{eqnarray} 
and

\begin{eqnarray} 
\rho_{03}=\sum_{B}\frac{g_{\rho_B}}{m_\rho^2}I_{3_B}\rho_B 
\label{isovector_field}
\end{eqnarray}  
 
where the total baryon density is 

\begin{eqnarray} 
\rho = \sum_B \rho_B =\frac{1}{2\pi^2} \sum_{B} \gamma_B \int^{k_B}_0 dk ~k^2  
\end{eqnarray}

$k_B$ being the Fermi momenta of a particular baryon species B and the spin degeneracy factor $\gamma$ is 2 in this case. 

 The scalar equation of motion in terms of $Y=x/x_0 = m_B^{\star}/m_B$ is given by

\begin{eqnarray} 
\sum_B \Biggl[(1-Y^2)-\frac{B}{C_{\omega_N}}(1-Y^2)^2+\frac{C}{C_{\omega_N}^2}(1-Y^2)^3  +2\frac{C_{\sigma_B}~C_{\omega_N}}{m_B^2 ~Y^4} \frac{\Bigl(\sum\limits_{B} g_{\omega_B} \rho_B\Bigr)^2}{\sum\limits_{B} {g_{\omega_B}}^2} \nonumber \\ - 2 \sum_B \frac{~C_{\sigma_B}~\rho_{SB}}{m_B~ Y} \Biggr]=0
\label{scalar_field}
\end{eqnarray}

where, the scalar density $\rho_{SB}$ of each baryon is
 
\begin{eqnarray} 
\rho_{SB}=\frac{\gamma}{2 \pi^2} \int^{k_B}_0 dk ~k^2 \frac{m_B^*}{\sqrt{k^2 + {m^{*}_{B}}^2}}
\end{eqnarray}
 
 As the momentum (density) increases within the core of NSs, the nucleon chemical potential reaches the rest mass state of the hyperons, they latter appearing in dense NSM. Similarly muons also appear at the expense of the electrons. The $\Delta$s are formed in NSM as resonance states via the strong interactions \cite{Sen}.
 
 
 The formation of these baryons are controlled by the charge neutrality and chemical potential conditions in order to obtain a stable charge neutral NS configuration at chemical equilibrium. We then take these baryons at equal footing with the nucleons.  

 On the basis of the above theory, the energy density ($\varepsilon$) and pressure ($P$) are calculated as 

\begin{eqnarray} 
\varepsilon &=& \frac{m_B^2}{8~C_{\sigma_B}}(1-Y^2)^2 - \frac{m_B^2 B}{12~C_{\omega_N}C_{\sigma_B}}(1-Y^2)^3
+\frac{C m_B^2}{16 ~C_{\omega_N}^2~ C_{\sigma_B}}(1-Y^2)^4 \nonumber \\ & + & \frac{1}{2Y^2}C_{\omega_N} \frac {\Bigl(\sum\limits_{B} g_{\omega_B} \rho_B\Bigr)^2}{\sum\limits_{B} {g_{\omega_B}}^2} + \frac{1}{2}~m_\rho^2 ~\rho_{03}^2 + \frac{1}{\pi^2} \sum_{B} \gamma_B \int_{0}^{k_B} k^2 \sqrt{(k^2+{m_B^*}^2)} ~dk \nonumber \\
&+& \frac{\gamma}{2\pi^2} \sum_{\lambda= e,\mu^-} \int_{0}^{k_\lambda} k^2 \sqrt{(k^2+{m_\lambda}^2)}~ dk 
\protect\label{EoS1}
\end{eqnarray}

\begin{eqnarray}         
P =&& -\frac{m_B^2}{8~C_{\sigma_B}}(1-Y^2)^2+\frac{m_B^2 B}{12~C_{\omega_N}~C_{\sigma_B}}(1-Y^2)^3
-\frac{C~ m_B^2}{16~ C_{\omega_N}^2 C_{\sigma_B}}(1-Y^2)^4 \nonumber \\
&&+\frac{1}{2Y^2}~C_{\omega_N} \frac {\Bigl(\sum\limits_{B} g_{\omega_B} \rho_B\Bigr)^2}{\sum\limits_{B} {g_{\omega_B}}^2}  
+ \frac{1}{2}~m_\rho^2 ~\rho_{03}^2 + \frac{1}{3\pi^2} \sum_{B} \gamma_B \int_{0}^{k_B} \frac{k^4}{ \sqrt{(k^2+{m_B^*}^2)}}~ dk \nonumber \\
&&+ \frac{\gamma}{6\pi^2} \sum_{\lambda= e,\mu^-} \int_{0}^{k_\lambda} \frac{k^4}{ \sqrt{(k^2+{m_\lambda}^2)}}~ dk
\protect\label{EoS2} 
\end{eqnarray} 

where $C_{i_B}=(g_{i_B}/m_i)^2$ are the scaled couplings with $i = \sigma, \omega, \rho$ and $C_{\omega_N}=1/{x_0^2}$.

\subsubsection{The model parameter\\}

 By fixing the properties of symmetric nuclear matter (SNM), the model parameter set is obtained self-consistently with relativistic MFT \cite{TKJ,Sen,Sen2,Sen3,param1,param2} at T $= 0$. The details of the procedure is available in \cite{TKJ} from which the parameter set for the present work is chosen. It is listed below in table \ref{table-1}, along with the saturation properties. 

\begin{table*}[ht!]
\centering
\tbl{Parameter set of the nuclear matter model considered for the present work (adopted from \cite{TKJ,Sen,Sen2,Sen3}). $C_{\sigma_N}$, $C_{\omega_N}$ and $C_{\rho_N}$ are the scalar, vector and iso-vector couplings, respectively. $B$ and $C$ are the higher order couplings of the scalar field. $m_{\sigma}$ is the scalar meson mass. The saturation properties such as binding energy per nucleon $B/A$, nucleon effective mass $m^{\star}_N$/$m_N$, the symmetry energy coefficient $J$, slope parameter ($L_0$) and nuclear incompressibility ($K$) defined at saturation density $\rho_0$ are also displayed.}
{{
\setlength{\tabcolsep}{15.5pt}
\begin{tabular}{ccccccccccccc}
\hline
\hline
\multicolumn{1}{c}{$C_{\sigma_N}$}&
\multicolumn{1}{c}{$C_{\omega_N}$} &
\multicolumn{1}{c}{$C_{\rho_N}$} &
\multicolumn{1}{c}{$B/m^2$} &
\multicolumn{1}{c}{$C/m^4$} &
\multicolumn{1}{c}{$m_{\sigma}$}\\
\multicolumn{1}{c}{($\rm{fm^2}$)} &
\multicolumn{1}{c}{($\rm{fm^2}$)} &
\multicolumn{1}{c}{($\rm{fm^2}$)} &
\multicolumn{1}{c}{($\rm{fm^2}$)} &
\multicolumn{1}{c}{($\rm{fm^2}$)} &
\multicolumn{1}{c}{(MeV)} \\
\hline
6.772  &1.995  & 5.285 &-4.274   &0.292  &510 \\
\hline
\hline
\multicolumn{1}{c}{$m^{\star}_N$/$m_N$}&
\multicolumn{1}{c}{$K$} & 
\multicolumn{1}{c}{$B/A$} &
\multicolumn{1}{c}{$J$} &
\multicolumn{1}{c}{$L_0$} &
\multicolumn{1}{c}{$\rho_0$} \\
\multicolumn{1}{c}{} &
\multicolumn{1}{c}{(MeV)} &
\multicolumn{1}{c}{(MeV)} &
\multicolumn{1}{c}{(MeV)} &
\multicolumn{1}{c}{(MeV)} &
\multicolumn{1}{c}{($\rm{fm^{-3}}$)} \\
\hline
0.85  &303  &-16.3   &32  &87  &0.153 \\
\hline
\hline
\end{tabular}
}}
\protect\label{table-1}
\end{table*}
 
The values nuclear incompressibility ($K = 303$~ MeV) and symmetry energy coefficient ($J = 32$~MeV) obtained for the model are consistent with that prescribed by \cite{k1,k2,Stone2,j1}. The slope parameter for the present model ($L_0 = 87$~MeV), although a bit larger than that suggested by \cite{l1}, is within the range  $L_0 = (25 - 115)$~MeV as suggested by \cite{rmf1}. Moreover, recently \cite{Fattoyev,ZhenYuZhu} have shown from the co-relation between the symmetry energy and tidal deformability and radius $R_{1.4}$ of NS, the value of $L_0$ can be upto $\sim$ 80 MeV, which is quite comparable with that obtained with the present model. The  binding energy per particle ($B/A = -16.3$~MeV) for SNM at the saturation density ($\rho_0 = 0.153$~$\rm{fm^{-3}}$), obtained with the model, fall within the range of values for the same \cite{rmf1}. The EoS, yielded by this parameter set (shown in table \ref{table-1}) for SNM and pure neutron matter (PNM) also passes through the heavy-ion collision data \cite{hic} as shown in \cite{TKJ}. However, owing to the high value of nucleon effective mass ($m_N^*=0.85~ m_N$) yielded by the present model, compared to other RMF models \cite{rmf,rmf1}, the obtained EoS softens at high density \cite{TKJ4,TKJ} and hence passes through the soft band of heavy-ion collision data \cite{TKJ}. The softening is thus expected to be becomes more pronounced with the inclusion of the hyperons and the $\Delta$s in the present work.

 This model along with the same parameter set (as given in table \ref{table-1}) has been successfully applied to study the nuclear matter properties at finite temperature \cite{TKJ4} and the properties of NSs with hyperon rich matter in both static \cite{TKJ2} and rotational configurations \cite{TKJ3} considering the potential depth of only the $\Lambda$ hyperon. Recently it has also been used to study the NS properties including only the $\Delta$ baryons in NSM and to understand the properties of hybrid stars in presence of the $\Delta$s in its hadronic phase along with the nucleons \cite{Sen}. The model parameters are well constrained and related to the VEV of the scalar field \cite{TKJ,Sahu,Sen,Sen2,Sen3}. The model presents very few free parameters to adjust the saturation properties. Thus the parameter set and the model adopted for the present work are amply tested and consistent with the recent experimental and empirical estimates of nuclear saturation properties and the heavy-ion collision data. 
 
\subsubsection{Hyperon and Delta coupling constants\\}
\label{Coupling}

 For the hyperons, the value of scalar coupling constants $x_{\sigma_H}={g_{\sigma_H}}/{g_{\sigma_N}}$ are chosen consistent to the limit ($x_{\sigma_H} \leq 0.72$) specified by \cite{Glen,Glen5,Rufa}. The vector couplings $x_{\omega_H}={g_{\omega_H}}/{g_{\omega_N}}$ are calculated reproducing the potential depths of the individual hyperon species ($(B/A)_H|_{\rho_0}$ = -28 MeV for $\Lambda$, +30 MeV for $\Sigma$ and -18 MeV for $\Xi$ \cite{Schaffner-Bielich,Sulaksono,be}) in SNM, using relation \ref{be} \cite{Glen,Glen5,Glen6,Arumugam,TKJ2,TKJ3} 

\begin{eqnarray}         
(B/A)_H\biggr|_{\rho_0} = x_{\omega_H} ~ g_{\omega_N} ~\omega_0 - x_{\sigma_H}~ g_{\sigma_N} ~\sigma_0
\protect\label{be}
\end{eqnarray}         

Here $x_{\rho_H}$ and $x_{\omega_H}$ are taken to be equal as both $\rho$ and $\omega$ mesons have almost similar mass and also both are responsible for the generation of short range repulsive forces.

 The $\Delta$-meson couplings, on the other hand, are still inconclusive. Numerous works have suggested different coupling schemes for this purpose. A detailed discussion on the uncertainties regarding the $\Delta$-meson couplings available in literature can be found in \cite{Sen}. At present there is no concrete values for the $\Delta$-meson couplings from experimental perspectives since the potential depth for the $\Delta$s is largely unknown. Refs. \cite{Riek,DragoPRC} with the references therein predict a shallow and attractive range for the $\Delta$ potential -30 MeV + $V_N$ $\leq V_{\Delta} \leq V_N$ in normal nuclear matter. Refs. \cite{Kolomeitsev,Maslov} on the other hand suggest the range of $\Delta$ potential to be $V_{\Delta}=-(50-100)$ MeV. It is therefore one can rely only on the theoretical predictions for the scalar and vector $\Delta$-meson couplings $x_{\sigma\Delta}={g_{\sigma\Delta}}/{g_{\sigma_N}}$ and $x_{\omega\Delta}={g_{\omega\Delta}}/{g_{\omega_N}}$. From the finite density QCD sum-rule calculations, \cite{qcd1} suggests larger scalar $\Delta$ coupling than those for the nucleons while the corresponding vector coupling for $\Delta$s to be two times smaller than those of the nucleons. On the other hand \cite{qcd,Lavagno,del1} have prescribed a range for the scalar and vector $\Delta$ couplings based on the three criteria : (i) the second minimum of the energy per baryon must lie above the saturation energy of normal nuclear matter, (ii) there are no $\Delta$ isobars present at the saturation density and (iii) the scalar field is more or same attractive while the vector potential is less or same repulsive for $\Delta$s compared to that of nucleons. The prescriptions of \cite{qcd1} based on QCD sum-rule predictions, do not match with the estimates suggested by \cite{qcd,Lavagno,del1} for the choice of scalar and vector couplings for the $\Delta$s, satisfying the above three requirements. It is therefore I abide by the range suggested by \cite{qcd,Lavagno,del1} in the form of a triangle for the scalar and vector $\Delta$ couplings. Due to lack of any concrete suggestion available in literature for the isovector $\Delta$ coupling  both from theoretical and experimental perspectives, moderate value of $x_{\rho\Delta}$ is chosen in this work.

 Along with the obtained EoS for the core, the well known BPS EoS \cite{BPS} is used to account for the crust part of the NS, having a much low density compared to the core. 
 
\subsection{Neutron Star Structure \& Properties in General Relativistic and parameterized hydrostatic equilibrium conditions}
\label{TOV-pTOV}

 On the basis of general relativistic (GR) theory, Tolman \cite{tov} and Oppenheimer \& Volkoff \cite{tov1} put forward the theory of hydrodynamic equilibrium between gravity and internal pressure of a star. The star in this case is considered to be undeformed sphere of perfect fluid in static condition. Under such approximations the Tolman-Oppenheimer-Volkoff (TOV) equations (eq. \ref{tov1} and \ref{tov2}) were derived to provide the global properties of NSs like the gravitational and baryonic masses, radius and central energy density.
 
\begin{eqnarray}
\frac{dM}{dr}= 4\pi r^2 \varepsilon
\protect\label{tov1}
\end{eqnarray} 

and

\begin{eqnarray}
\frac{dP}{dr}=-\frac{GM(r)\varepsilon}{r^2}\frac{\Big(1+\frac{P}{\varepsilon}\Big)\Big(1+\frac{4\pi^2 r^3 P}{M(r)}\Big)}{\Big(1-\frac{2GM(r)}{r}\Big)}
\protect\label{tov2}
\end{eqnarray} 

with $G$ as the gravitational constant and $M(r)$ as the enclosed gravitational mass for a given choice of central energy density $(\varepsilon_c)$ and specified EoS. Here $c=1$. The surface of the star where the pressure vanishes gives the value of radius $r~(=R)$ of the star. The baryonic mass $M_B(r)$ can also be calculated as

\begin{eqnarray} 
M_B(r)=\int_{0}^{R} 4\pi r^2 ~\varepsilon~ m_B \left(1 - \frac{2GM}{r}\right)^{1/2} dr
\label{barmass}
\end{eqnarray}

where, $m_B$ is the mass of baryon.

\vspace{1cm}

 The parameterized TOV (PTOV) equations \cite{Velten} are given as

\begin{eqnarray}
\frac{dM}{dr}= 4\pi r^2 (\varepsilon + \tilde{\sigma} P)
\protect\label{modtov1}
\end{eqnarray} 

and

\begin{eqnarray}
\frac{dP}{dr}=-\frac{G(1+ \alpha)M(r)\varepsilon}{r^2}\frac{\Big(1+ \beta \frac{P}{\varepsilon}\Big)\Big(1+ \chi \frac{4\pi^2 r^3 P}{M(r)}\Big)}{\Big(1- \tilde{\gamma} \frac{2GM(r)}{r}\Big)}
\protect\label{modtov2}
\end{eqnarray}    

The five free parameters $\alpha$, $\beta$, $\chi$, $\tilde{\gamma}$ and $\tilde{\sigma}$ are independent of eachother and each have important physical significance. 
 
\begin{itemize}

\item $\alpha$ measures the net effective gravitation and is mostly important for f(R) gravity theories \cite{Brax}. It is taken to be 1/3 in such case. 
 
 It is clear from the general TOV equation (eq. \ref{tov2}) that the pressure have two major contributions \cite{Velten,Schwab} to
 
 i) the total mass density as $\Big(\varepsilon+P\Big)$ and 
 
 ii) the total gravitational mass $\Big(M(r)+4\pi^2 r^3 P\Big)$. Therefore the two separate effects of pressure are parameterized by 
 
\item $\beta$ for the inertial pressure that contribute to the total mass density which appears as a consequence of conservation of the energy-momentum tensor and 
 
\item $\chi$ parametrizes the gravitational mass of NS, controlled by pressure, giving rise to the self-gravity of the star \cite{Schwab}.
 
\item $\tilde{\gamma}$ is associated with the curvature contribution to the star, depending on its geometry. This a unique feature of GR configuration and $\tilde{\gamma}=0$ gives the Newtonian case for which the mass of NS becomes too high that the the causality limit \cite{Latt-Prak2007} is violated. This effect is also seen in \cite{Velten}.
 
\item $\tilde{\sigma}$ introduces the effect of gravity to mass function. For neo-Newtonian case, $\tilde{\sigma}= 3$ \cite{Oliveira2} but it is not a reasonable treatment for NSs. Therefore we keep $\tilde{\sigma}=0$ as in normal GR case. 
\end{itemize} 

 In normal GR case ($\alpha$,$\beta$,$\chi$,$\tilde{\gamma}$,$\tilde{\sigma}$) have configuration (0,1,1,1,0) to get back the general TOV eqs. \ref{tov1}, \ref{tov2} while the configuration ($\alpha$,$\beta$,$\chi$,$\tilde{\gamma}$,$\tilde{\sigma}$)=(0,0,0,0,0) leads to pure Newtonian case. 
 
 In the present work the two separate effects of pressure on the properties of massive NSs are highlighted by varying the two parameters $\beta$ and $\chi$ and keeping the others same as normal GR configuration. The configuration ($\beta$,$\chi$)=(0,0) does not correspond to the Newtonian case. This is because the effects of GR conditions are still rendered via $\tilde{\gamma}=1$ as it relates to the notion of curvature of the star, which is very pronounced in case of NSs. Under such circumstance (i.e, with $\tilde{\gamma}=1$), the different configurations of $\beta$ and $\chi$ signify modifications to normal GR conditions in terms of pressure. I vary the parameters $\beta$ and $\chi$ within the extent suggested by \cite{Velten,Schwab}. The conservation of energy–momentum tensor gives rise to the notion of inertial mass density, which in turn is proportional to the force that the fluid must experience in order to overcome gravitational collapse. It is well known that this force is greater for massive NSs. Thus the parameter $\beta$, coupling to the inertial mass density, actually measures the binding force of the fluid, which is more intense for massive NSs. Another important contribution of pressure is total gravitational mass.. The change in pressure $dP/dr$ is the equivalent force that accelerates the fluid away from its geodesic \cite{Schwab}. This gives rise to the concept of total gravitational mass. As this force becomes more prominent and the acceleration of the fluid enhances more for massive NSs, the parameter $\chi$ in such cases relates to the change in gravitating effects of pressure and the self gravity of the star for massive NSs.
 
 Though the variation of these parameters are physically justified in case of massive NSs, however, the extent to which they can be varied is still inconclusive. One possible way they may be constrained is in terms of the observed mass of massive pulsars like PSR J1614−2230 and PSR J0348+0432 within the framework of a chosen model. This is shown in one of our recent works \cite{Sen3} where we have presented a possible bound on $\beta$ and $\chi$ with respect to the masses of these pulsars with our hyperonic EoS. The necessary deviation from normal GR conditions in terms of $\beta$ and $\chi$ to achieve high NS mass configurations are presented in Ref. \cite{Sen3} with the same effective chiral model. With the availability of a vast number of models and EoS in present literature, the variations of $\beta$ and $\chi$ in order to achieve high NS mass configurations may thus depend on particular model and the EoS considered. However, considering our effective chiral model and including the hyperons and $\Delta$s, the importance of incorporating these modifications to normal GR via $\beta$ and $\chi$ are presented in the next section.

\section{Result and Discussions}
\subsection{Neutron Star matter including the baryon octet and $\Delta$ quartet}

 The scalar coupling of the hyperons is fixed as $x_{\sigma_H}=0.7$, which is within the bound on $x_{\sigma_H}$ \cite{Glen,Glen5,Rufa} and the corresponding values of $x_{\omega_H}$ are calculated according to eq. \ref{be} as discussed in section \ref{Coupling}. The scalar and vector delta couplings are chosen according to the finite density QCD sum-rule calculations \cite{qcd,qcd1}. It is suggested by QCD sum-rule calculations that $x_{\sigma_\Delta} \geq 1$ and $x_{\omega_\Delta} \leq 1$. As mentioned earlier that there is no theoretical or experimental prescription for the isovector delta coupling in literature. But it is seen from \cite{prc92,Sen} that the formation of the $\Delta$s and the NS properties are very sensitive to the $x_{\rho_\Delta}$ couplings. A higher value of $x_{\rho_\Delta}$ yields comparatively massive NS configurations \cite{prc92,Sen}. However, as mentioned in our earlier work \cite{Sen} $\Delta$s are not obtained when $x_{\rho_\Delta} > 1$. Therefore for the present work the delta couplings are chosen as $(x_{\sigma_\Delta},x_{\omega_\Delta},x_{\rho_\Delta}) = (1.35,1.0,1.0)$. It is to be noted that this set of delta couplings has been used in a previous analysis for formation of $\Delta$s and the resultant NS properties \cite{Sen}. For this set the delta potential obtained in SNM is -110 MeV, which is quite larger than the range suggested for the same by \cite{DragoPRC}. The obtained value of potential is, however, consistent with the range prescribed by \cite{Kolomeitsev,Maslov}. It is seen from \cite{Sen} that with this coupling set the second minima of the energy per baryon lie well above the saturation energy of normal nuclear matter (-16.3 MeV). For the chosen coupling set the second minima lie approximately at 8 MeV (2.5 $\rho_0$) \cite{Sen}.
 
\begin{figure}[!ht]
\centering
\includegraphics[scale=1.0]{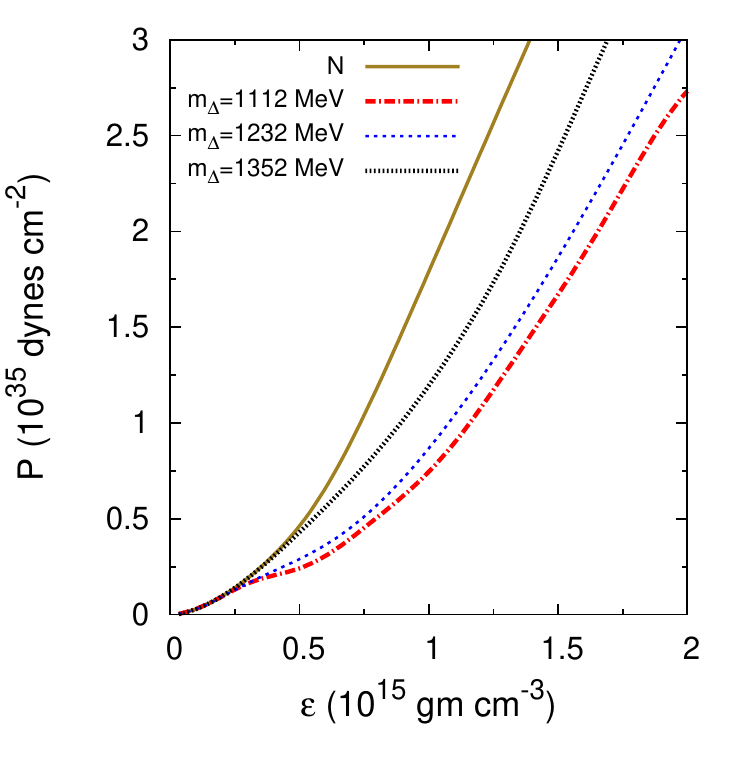}
\caption{\label{eosall} Equation of State ($\varepsilon ~vs.~ P$) for neutron star matter including different baryons and leptons with different masses of $\Delta$s.}
\end{figure}

  The EoS is obtained including the $\Delta$s with varied mass $m_{\Delta}$ in the presence of hyperons and compared with the EoS for pure nucleon matter (N) in fig. \ref{eosall}.
  
As seen from the figure, the EoS softens largely from the pure nucleon matter case (N) due to formation of the hyperons and the $\Delta$s. The softening is most prominent when $m_{\Delta}=1112$ MeV (minimum). The EoS stiffens with the increase of $m_{\Delta}$. This is because the formation of less massive particles are favored in NSM with consequent softening of the EoS. Therefore it is expected that the resultant maximum mass of the NS will be reduced considerably compared to that of the pure nucleon matter case. This will be discussed subsequently.

\begin{figure}[!ht]
\centering
\includegraphics[scale=0.6]{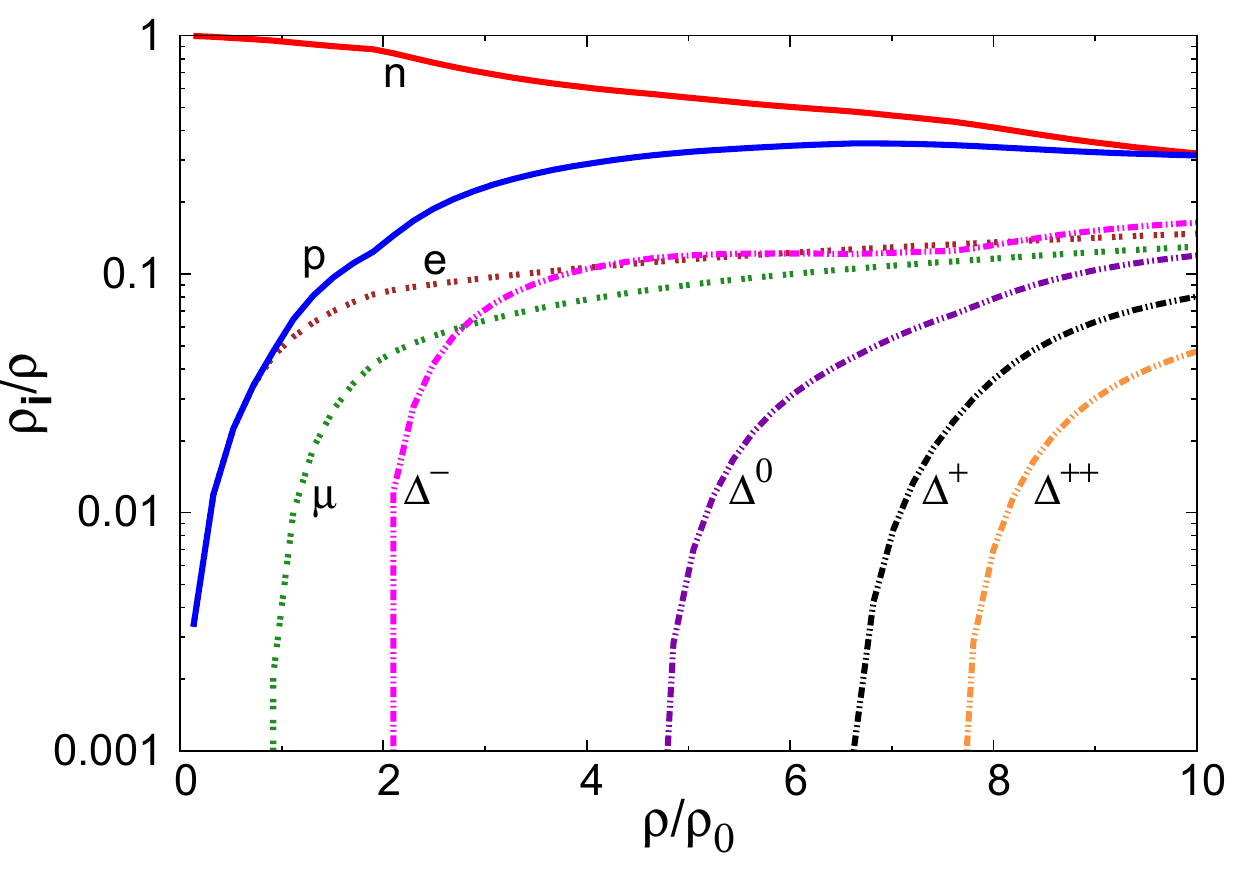}
\caption{\label{pf1112} Relative particle fraction of different baryons and leptons in neutron star matter for $m_{\Delta}=1112$ MeV.}
\end{figure}

\begin{figure}[!ht]
\centering
\includegraphics[scale=0.6]{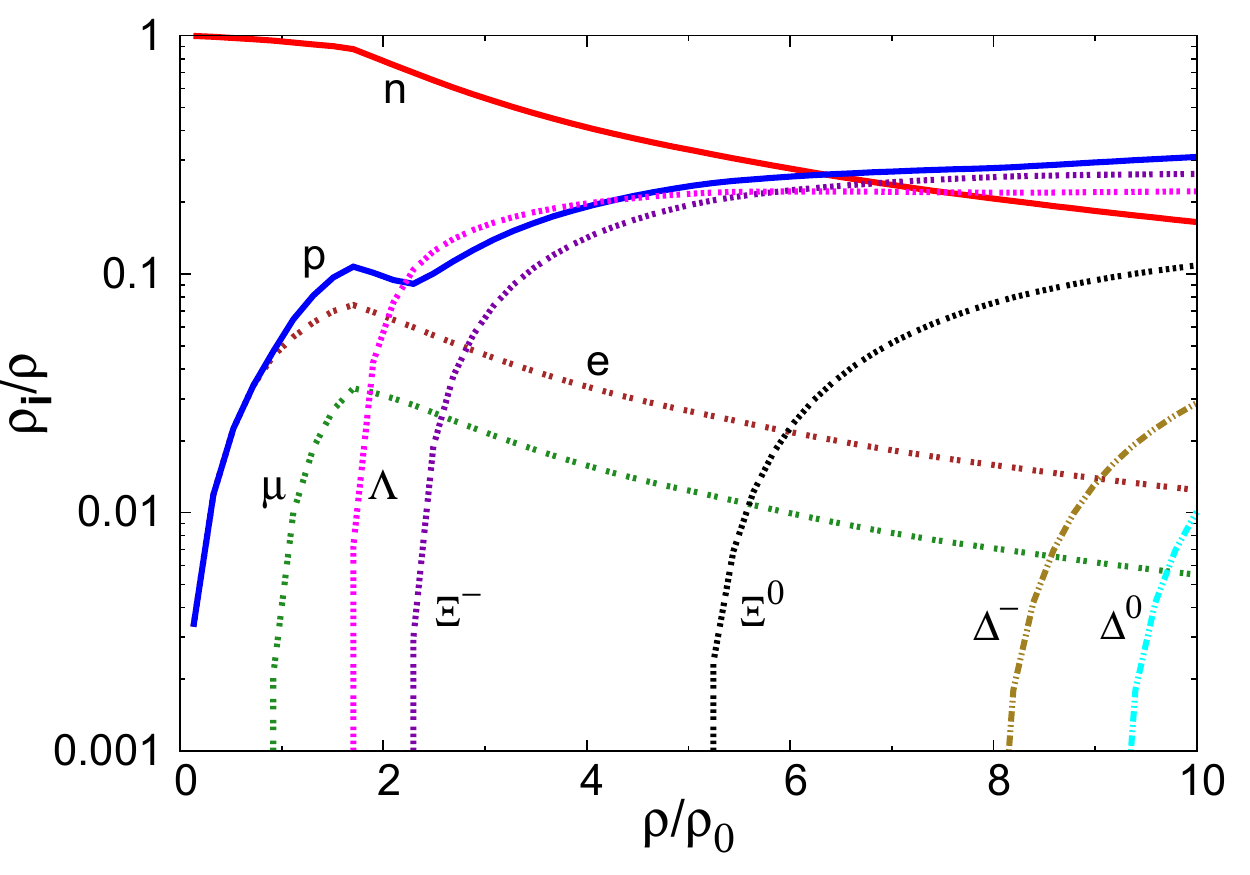}
\caption{\label{pf1232} Relative particle fraction of different baryons and leptons in neutron star matter for $m_{\Delta}=1232$ MeV.}
\end{figure}

\begin{figure}[!ht]
\centering
\includegraphics[scale=0.65]{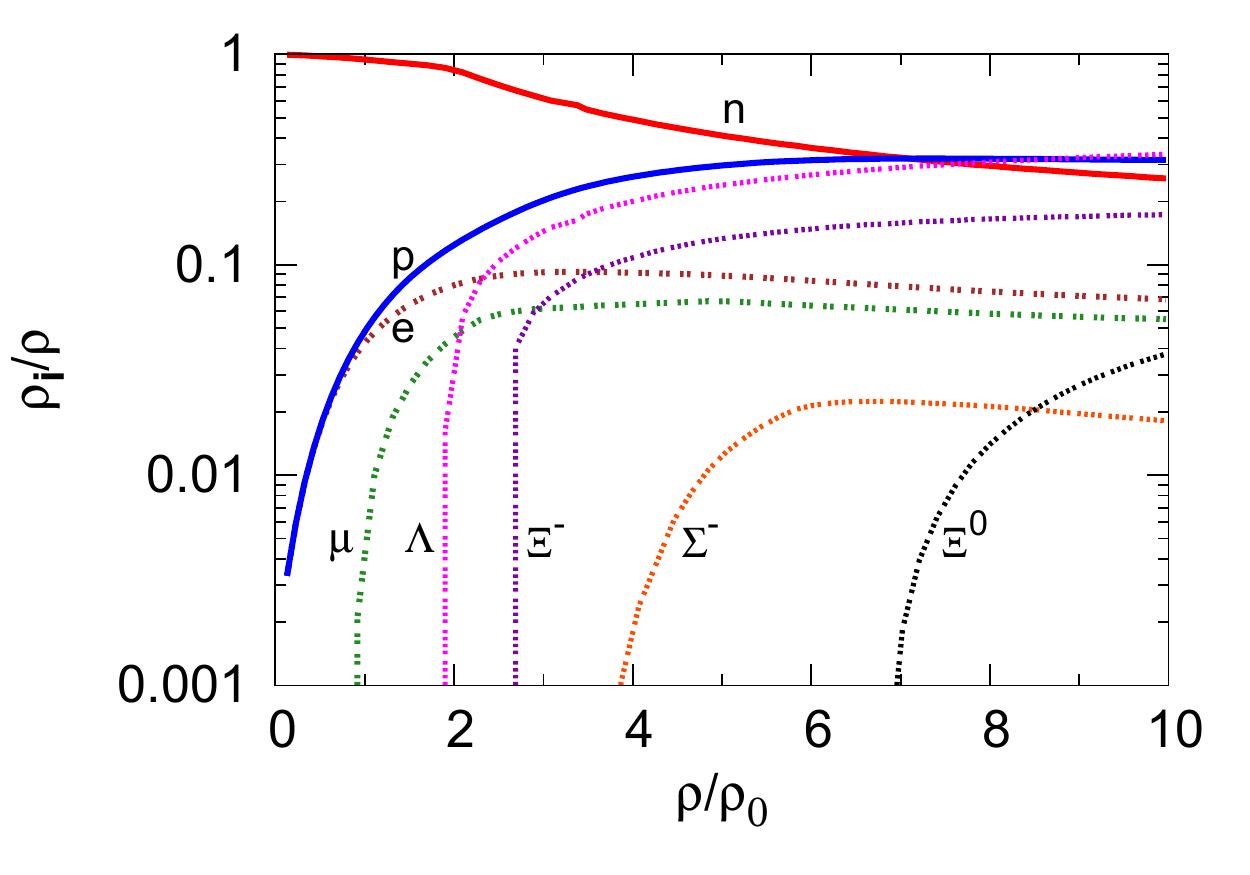}
\caption{\label{pf1352} Relative particle fraction of different baryons and leptons in neutron star matter for $m_{\Delta}=1352$ MeV.}
\end{figure}

  In figs. \ref{pf1112},\ref{pf1232} and \ref{pf1352} the relative population fraction (${\rho_i}/{\rho}$) of different baryons and leptons as a function of normalized baryon density (${\rho}/{\rho_0}$) for different values of $\Delta$ mass.

 It is seen that at the expense of neutrons, hyperons and $\Delta$s are formed in NSM in considerable amount depending on the value of $m_{\Delta}$. As it is well known that the formation of light and negatively charged particles are most favored in NSM, all the $\Delta$s populate in NSM suppressing the formation of all the hyperons when the $\Delta$ mass is taken to be least ($m_{\Delta}=1112$ MeV) (fig. \ref{pf1112}). In this case the $\Delta^-$, being charge favored, appears first and as early as at 2.1 $\rho_0$. The next to appear is the charge neutral $\Delta^0$ at 4.9 $\rho_0$. The positively charged $\Delta^+$ appears next at 6.7 $\rho_0$ followed by $\Delta^{++}$ at 7.8 $\rho_0$. It is noteworthy that the concentration of the negatively charged leptons in this case is quite high to balance the net positive charge in NSM due to the formation of $\Delta^+$ and $\Delta^{++}$.
 
 As the delta mass is increased to a moderate value of $m_{\Delta}=1232$ MeV it is found that the $\Delta$s are no longer the most populated particles in NSM (fig. \ref{mr1232}). Their formation is rather quite delayed and reduced in concentration. In this case $\Lambda$ appears first at 1.8 $\rho_0$. The next to form is $\Xi^-$ at density 2.4 $\rho_0$, followed by $\Xi^0$ at 5.1 $\rho_0$. Among the deltas, only $\Delta^-$ and $\Delta^0$ are formed at 8.0 $\rho_0$ and 9.3 $\rho_0$. Consistent with the results of many works like \cite{Glen6}, the formation of $\Sigma$s are totally restricted in NSM because of their repulsive potential depth in nuclear matter. Both $\Xi$ and $\Delta$s, despite of being heavier than the $\Sigma$s, are formed in NSM instead of the $\Sigma$s. This is because unlike the $\Sigma$s the $\Xi$ and $\Delta$s suffer much less repulsion in NSM due to their negative potential depths. $\Sigma$s on the other hand experience higher repulsion owing to their positive potential depth (+30 MeV). It is also seen from fig. \ref{mr1232} that with the increase in value of $m_{\Delta}$ compared to that in fig. \ref{pf1112}, the $\Delta$s no longer remain the most favored and populated particle in NSM. $\Delta^+$ and $\Delta^{++}$ are not formed due to rapid exhaustion of neutrons while $\Delta^-$ and $\Delta^0$ are formed very late with much reduced concentration. Interestingly, contrary to the previous one, deleptonization occurs very fast in this case due to excess generation of negative charge by the high concentration of $\Xi^-$ and $\Delta^-$ to some extent.
 
 With further increase in mass of $\Delta$s to $m_{\Delta}=1352$ MeV, their formation in NSM is totally suppressed by the formation of hyperons (fig. \ref{mr1352}). Similar to that of fig. \ref{mr1232}, $\Lambda$ appears first at 1.9 $\rho_0$ and $\Xi^-$ is formed next at 2.7 $\rho_0$. $\Sigma^-$ then appears at 3.9 $\rho_0$ followed by $\Xi^0$ at 7.0 $\rho_0$. It is seen that the heavier mass of the $\Delta$s are totally unfavorable for them to appear in NSM. On the other hand $\Sigma$s, suffering from high repulsion, their formation still remains quite restricted in NSM. However, due to suppression of the massive $\Delta$s, only $\Sigma^-$ appears as it is charge favored. But despite of being more massive than the $\Xi$s, it appears late and much after $\Xi^-$. The formation of $\Sigma^0$ and $\Sigma^+$ remains totally subdued by their positive potential depth. It is interesting to note that $\Sigma^-$, though formed late, unlike other hyperons its concentration diminishes after a certain value of density (6.7 $\rho_0$) owing to the dominance of vector repulsion on the $\Sigma$s at high densities. There is moderate population of the negatively charged leptons compared to the previous cases in order to maintain the overall charge neutrality of NSM.

 Overall, it is quite interesting to note that with three different masses of the deltas, the formation of different baryons and leptons changes considerably. With low mass, the $\Delta$s have the privileged of being the only baryons formed with the nucleons in large amount (total 43\%). Moreover, all the $\Delta$s appear when their mass is considered to be minimum. The scenario, however, changes and it is seen that with a moderate increase in their mass, there is much delayed formation of only $\Delta^-$ and $\Delta^0$ with much reduced concentration (only 4\% in total). Further if they are considered to be more massive, the $\Delta$s are not at all formed in NSM due to rapid exhaustion of neutrons.
 
\subsection{Static Neutron Star properties in general hydrostatic equilibrium} 
 
 Subjecting the obtained EoSs for different $m_{\Delta}$ first to the general TOV equations \ref{tov1} and \ref{tov2}, the various static properties of NS like central energy density ($\varepsilon_c$), gravitational mass ($M$), baryonic mass ($M_B$), radius ($R$) and radius of canonical mass ($R_{1.4}$) and $R_{1.6}$ are calculated. They  are tabulated in table \ref{Stat.PropTOV} along with that obtained for the pure nucleon case (N).
 
\begin{table*}[ht!]
\centering
\tbl{Static neutron star properties for pure nucleon matter (N) and neutron star matter with hyperons (H) \& $\Delta$s for different delta mass ($m_{\Delta}$). The results from hydrostatic equilibrium conditions such as the central density of the star $\varepsilon_c$ ($\times 10^{15}$ g cm$^{-3}$), the maximum gravitational mass $M$ ($M_{\odot}$), maximum baryonic mass $M_B$ ($M_{\odot}$) radius $R$ (km), $R_{1.4}$ (km) and $R_{1.6}$ (km) are displayed.}
{{
\setlength{\tabcolsep}{10.0pt}
\begin{tabular}{cccccccccc}
\hline
\hline
\multicolumn{1}{c}{}&
\multicolumn{1}{c}{$m_{\Delta}$}&
\multicolumn{1}{c}{$\varepsilon_c$}&
\multicolumn{1}{c}{$M$}&
\multicolumn{1}{c}{$M_{B}$} &
\multicolumn{1}{c}{$R$} &
\multicolumn{1}{c}{$R_{1.4}$} &
\multicolumn{1}{c}{$R_{1.6}$} & \\
%
\multicolumn{1}{c}{} &
\multicolumn{1}{c}{(MeV)} &
\multicolumn{1}{c}{($\times 10^{15}$ g cm$^{-3}$)} &
\multicolumn{1}{c}{($M_{\odot}$)} &
\multicolumn{1}{c}{($M_{\odot}$)} &
\multicolumn{1}{c}{(km)} &
\multicolumn{1}{c}{(km)} &
\multicolumn{1}{c}{(km)} & \\
\hline
N            &-    &1.22 &2.10 &2.41 &12.2 &13.4 &13.3 \\
N+H+$\Delta$ &1112 &1.45 &1.65 &1.77 &10.7 &11.6 &11.0 \\
N+H+$\Delta$ &1232 &1.53 &1.69 &1.82 &10.4 &12.3 &11.2 \\
N+H+$\Delta$ &1352 &1.84 &1.76 &1.96 &11.2 &12.5 &12.4 \\

\hline
\hline
\end{tabular}
}}
\protect\label{Stat.PropTOV}
\end{table*}

\begin{figure}[!ht]
\begin{center}
\includegraphics[scale=0.8]{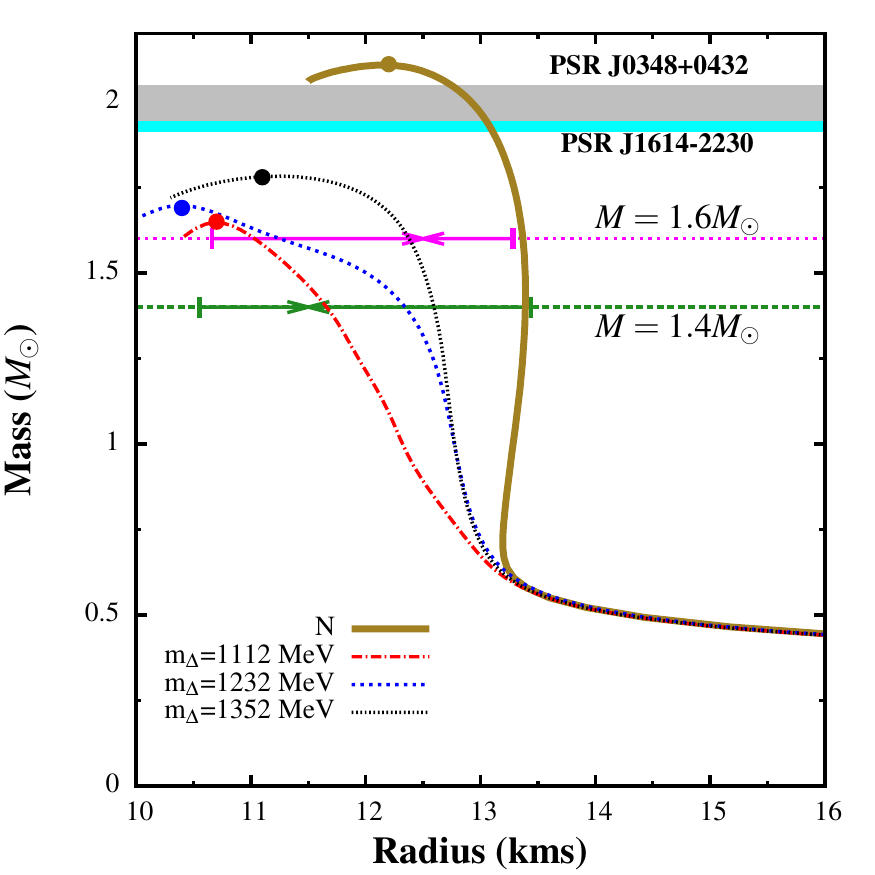}
\caption{\it Variation of mass with radius of neutron stars in static normal GR conditions for pure nucleons (N) and including hyperons and $\Delta$s for different $m_{\Delta}$. Observational limits imposed from high mass stars PSR J1614-2230 ($M = (1.928 \pm~ 0.017) M_{\odot}$) \cite{Fonseca} (cyan band) and PSR J0348+0432 ($M = (2.01 \pm 0.04) M_{\odot}$) \cite{Ant} (grey band) are also indicated. The points of maximum mass obtained are shown. The green horizontal line indicates the canonical mass ($M = 1.4~M_{\odot}$) while mass $M = 1.6~M_{\odot}$ is marked with magenta. Range of $R_{1.4}$ is marked according to \cite{Abbott,Fattoyev} while that of $R_{1.6}$ is marked according to \cite{Abbott} and \cite{Bauswein}.}
\protect\label{mrall}
\end{center}
\end{figure}

The mass-radius relationship with normal GR conditions for pure nucleonic matter (N) and NSM including hyperons and $\Delta$s are shown in fig. \ref{mrall} for different $m_{\Delta}$.

For the pure nucleonic case (N) the gravitational mass is 2.10 $M_{\odot}$ with corresponding radius 12.2 km.
As expected the gravitational mass reduces a lot due to formation of the hyperons and the $\Delta$s compared to the pure nucleonic case. The decrease is more prominent with the decrease in delta mass as the EoS softens gradually with the same (fig. \ref{eosall}). The maximum gravitational mass for delta masses $m_{\Delta}=$1112, 1232 and 1352 MeV are obtained to be 1.65 $M_{\odot}$, 1.69 $M_{\odot}$ and 1.76 $M_{\odot}$, respectively. The radii of 1.4 $M_{\odot}$ and 1.6 $M_{\odot}$ stars are consistent with the limits imposed from the results of GW170817 \cite{Abbott,Fattoyev,Bauswein} for pure nucleonic case (N) and also for matter including hyperons and $\Delta$s with all the values of $m_{\Delta}$. The central density is seen to increase with gravitational mass as the delta mass is increased. As a result of reduction in gravitational mass due the formation of these exotic particles (hyperons and $\Delta$s) the maximum mass constraint of $\sim 2 M_{\odot}$ from PSR J0348+0432 \cite{Ant} is not satisfied within the framework of GR. This leads to the well known "delta puzzle". 

\subsection{Static Neutron Star properties in parameterized hydrostatic equilibrium}

 Next the properties of NS are calculated using the PTOV equations \ref{modtov1} and \ref{modtov2} for NSM including hyperons and $\Delta$s with different delta masses. As discussed earlier that in this work the effects of pressure on the static properties are studied by varying the parameters $\beta$ and $\chi$ keeping the other parameters $\alpha$, $\tilde{\gamma}$ and $\tilde{\sigma}$ fixed to normal GR conditions. The results are displayed in table \ref{Stat.PropPTOV}
 
\begin{table*}[ht!]
\centering
\tbl{Static neutron star properties with respect to the variation in $\beta$ and $\chi$ considering neutron star matter with hyperons and $\Delta$s for different $m_{\Delta}$. Other parameters $\alpha$,$\tilde{\gamma}$ and $\tilde{\sigma}$ are fixed to normal GR conditions ($\alpha=0,\tilde{\gamma}=1,\tilde{\sigma}=0$). The results from parameterized TOV such as the central density of the star $\varepsilon_c$ (in $\rm{g~cm^{-3}}$), the mass $M$ (in $M_{\odot}$), the baryonic mass $M_B$ (in $M_{\odot}$) and the radius $R$ (in km), radius of canonical mass $R_{1.4}$ (in km) and $R_{1.6}$ (in km) are tabulated.}
{{
\setlength{\tabcolsep}{10.0pt}
\begin{tabular}{cccccccccc}
\hline
\hline
\multicolumn{1}{c}{$m_{\Delta}$}&
\multicolumn{1}{c}{$\beta$}&
\multicolumn{1}{c}{$\chi$} &
\multicolumn{1}{c}{$\varepsilon_c$} &
\multicolumn{1}{c}{$M$} &
\multicolumn{1}{c}{$M_{B}$} &
\multicolumn{1}{c}{$R$} &
\multicolumn{1}{c}{$R_{1.4}$} & 
\multicolumn{1}{c}{$R_{1.6}$} & \\
\multicolumn{1}{c}{(MeV)} &
\multicolumn{1}{c}{ } &
\multicolumn{1}{c}{ } &
\multicolumn{1}{c}{($\times 10^{15} \rm{g~cm^{-3}}$)} &
\multicolumn{1}{c}{($M_{\odot}$)} &
\multicolumn{1}{c}{($M_{\odot}$)}&
\multicolumn{1}{c}{$(km)$} &
\multicolumn{1}{c}{$(km)$} & 
\multicolumn{1}{c}{$(km)$} & \\
\hline
1112 & 0 & 0 &1.45 &1.98 &2.16 &10.5 &12.0 &11.7\\
     & 1 & 0 &1.44 &1.92 &2.11 &10.4 &11.8 &11.5\\
     & 0 & 1 &1.44 &1.87 &2.05 &10.4 &11.8 &11.4\\
     & 1 & 1 &1.45 &1.65 &1.77 &10.7 &11.6 &11.0\\
\hline
1232 & 0 & 0 &1.54 &2.03 &2.23 &10.4 &12.7 &12.5\\
     & 1 & 0 &1.52 &1.97 &2.14 &10.3 &12.6 &12.4\\
     & 0 & 1 &1.52 &1.87 &2.05 &10.3 &12.6 &12.2\\
     & 1 & 1 &1.53 &1.69 &1.82 &10.4 &12.3 &11.2\\
\hline
1352 & 0 & 0 &1.84 &2.13 &2.25 &11.6 &13.2 &13.1\\
     & 1 & 0 &1.83 &2.08 &2.17 &11.3 &13.0 &12.9\\
     & 0 & 1 &1.84 &1.97 &2.07 &11.2 &12.8 &12.7\\
     & 1 & 1 &1.84 &1.76 &1.96 &11.2 &12.5 &12.4\\      
\hline
\hline
\protect\label{Stat.PropPTOV}
\end{tabular}
}}
\end{table*}

\begin{figure}[!ht]
\begin{center}
\includegraphics[scale=0.8]{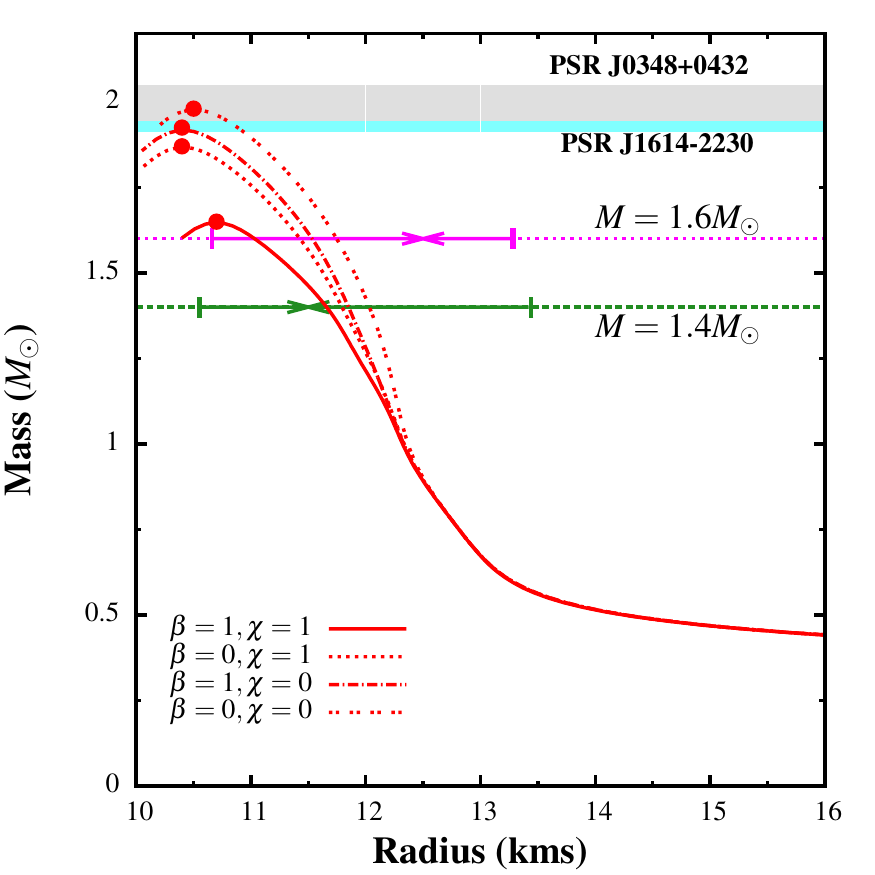}
\caption{\it Variation of mass with radius for different configurations of $\beta$ and $\chi$ for possible existence of different baryons and leptons with $m_{\Delta}=1112$ MeV. The solid curve shows general TOV solution. Observational limits imposed from high mass stars PSR J1614-2230 ($M = (1.928 \pm~ 0.017) M_{\odot}$) \cite{Fonseca} (cyan band) and PSR J0348+0432 ($M = (2.01 \pm 0.04) M_{\odot}$) \cite{Ant} (grey band) are also indicated. The points of maximum mass obtained are shown. The green horizontal line indicates the canonical mass ($M = 1.4~M_{\odot}$) while mass $M = 1.6~M_{\odot}$ is marked with magenta. Range of $R_{1.4}$ is marked according to \cite{Abbott,Fattoyev} while that of $R_{1.6}$ is marked according to \cite{Abbott} and \cite{Bauswein}.}
\protect\label{mr1112}
\end{center}
\end{figure}

\begin{figure}[!ht]
\begin{center}
\includegraphics[scale=0.9]{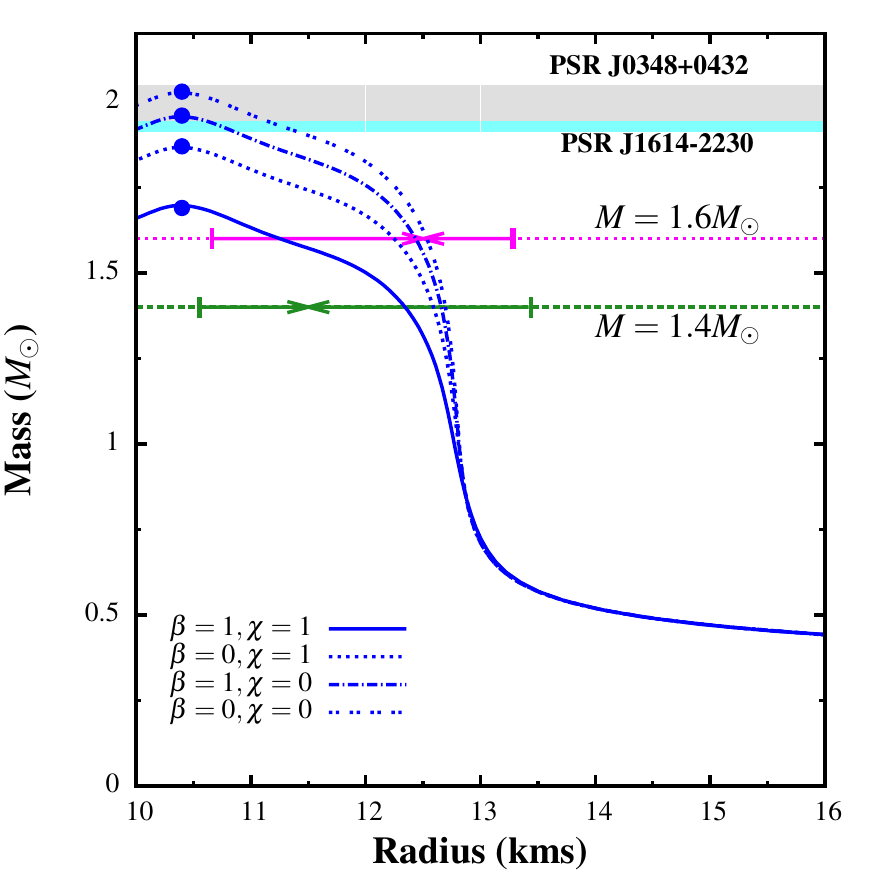}
\caption{\it Same as fig. \ref{mr1112} but for $m_{\Delta}=1232$ MeV.}
\protect\label{mr1232}
\end{center}
\end{figure}

\begin{figure}[!ht]
\begin{center}
\includegraphics[scale=0.87]{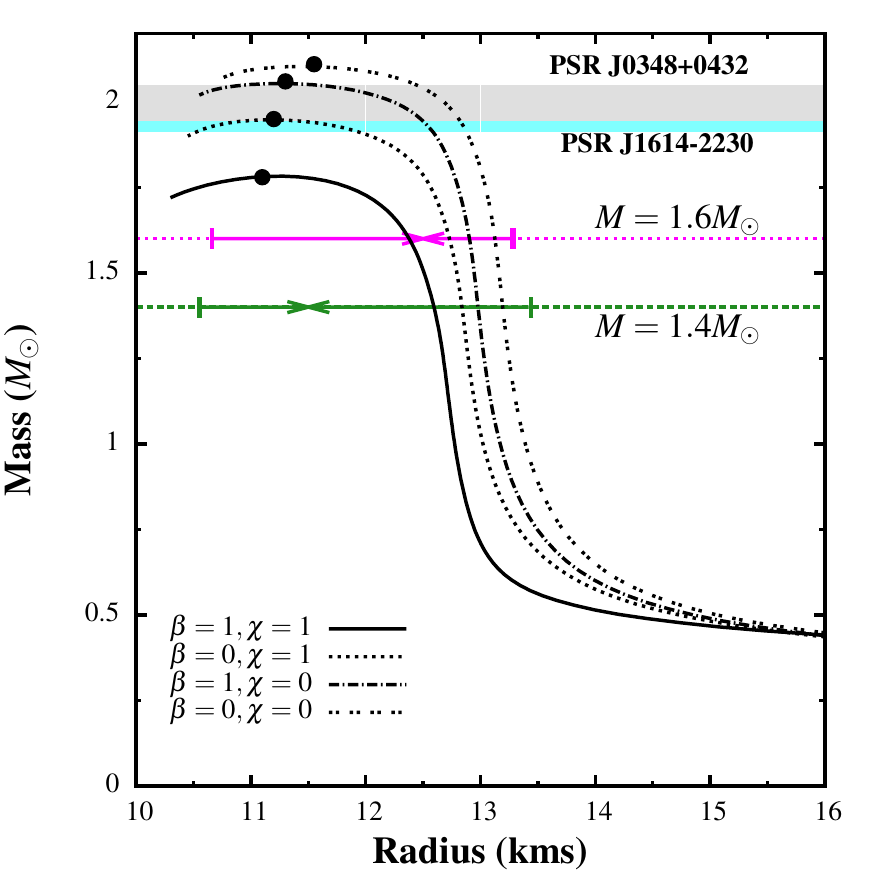}
\caption{\it Same as fig. \ref{mr1112} but for $m_{\Delta}=1352$ MeV.}
\protect\label{mr1352}
\end{center}
\end{figure}

Figs. \ref{mr1112}, \ref{mr1232} and \ref{mr1352} show the variation of gravitational mass with radius as $\beta$ and $\chi$ are varied with the other parameters fixed at normal GR conditions for NSM including hyperons and $\Delta$s for $m_{\Delta}=$1112, 1232 and 1352 MeV, respectively. 

 For every mass of the deltas, it seen from figs. \ref{mr1112}, \ref{pf1232} and \ref{pf1352} that the gravitational mass is minimum for normal GR case (($\beta$,$\chi$)=(1,1)). As the normal GR conditions are modified in terms of pressure via $\beta$ and $\chi$, it is seen that there is considerable increase in the gravitational mass of the NSs with moderate change in radius. For all the three values of delta mass the configuration ($\beta$,$\chi$)=(0,0) yields the maximum mass of the NSs. It is seen that the maximum gravitational mass decreases as the value of inertial pressure contributing to total mass density is increased via $\beta$. However, the effect of pressure contributing to total gravitational mass (self gravity) via $\chi$ on NS mass is just the opposite i.e. maximum gravitational mass increases with decreasing value of i.e. maximum gravitational mass increases with decreasing value of $\chi$. For example for $m_{\Delta}=1112$ MeV, keeping $\chi$ fixed at 0, the gravitational mass is 1.92 $M_{\odot}$ for $\beta=1$ and 1.98 $M_{\odot}$ for $\beta=0$. Same results are observed if variation of $\beta$ are observed with $\chi$ fixed at 1. On the other hand fixing $\beta=0$, the maximum mass is 1.98 $M_{\odot}$ for $\chi=0$ and 1.87 $M_{\odot}$ for $\chi=1$. Similar behavior is noticed when the value of $\chi$ is varied with $\beta$ fixed at 1. The results are consistent with that of of \cite{Velten}. The variation of these parameters brings moderate changes to the radius with feeble variation in central density.
 
 With the modified effects of pressure the ($M = (2.01 \pm 0.04) M_{\odot}$) constraint from PSR J0348+0432 \cite{Ant} is successfully satisfied for all the masses of delta, thereby resolving the delta puzzle. For $m_{\Delta}=1112$ MeV, this constraint is satisfied by the configuration ($\beta$,$\chi$)=(0,0) while for $m_{\Delta}=1232$ MeV it is satisfied by the configurations (0,0) and (1,0). For $m_{\Delta}=1352$ MeV, it is satisfied by (0,0), (1,0) and (0,1). With all the modified configurations of $\beta$ and $\chi$, the values of $R_{1.4}$ and $R_{1.6}$ are found to increase from that obtained in normal GR condition. The values of $R_{1.4}$ and $R_{1.6}$, obtained with all the configurations of $\beta$ and $\chi$ for each of the delta masses, fall within the range estimated from GW170817 data \cite{Abbott,Fattoyev,Bauswein}. The values of $R_{1.4}$ and $R_{1.6}$ are approximately obtained to be 11.8 km and 11.4 km, respectively for $m_{\Delta}=1112$ MeV. For $m_{\Delta}=1232$ MeV, the values are $R_{1.4}\approx12.5$ km and $R_{1.6}\approx12.1$ km. With $m_{\Delta}=1352$ MeV, $R_{1.4}\approx12.9$ km and $R_{1.6}\approx12.8$ km.
 
\begin{figure}[!ht]
\centering
\includegraphics[scale=0.83]{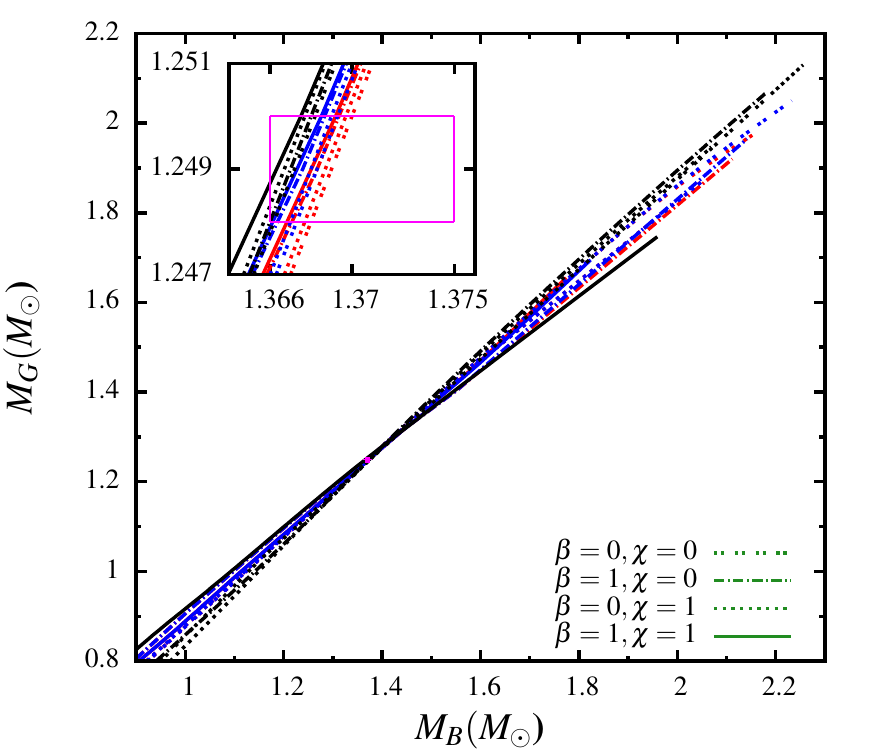}
\caption{\it Baryonic mass ($M_B$) versus gravitational mass ($M_G$) for static neutron star for different configurations of $\beta$ and $\chi$ for possible existence of different baryons and leptons. Red curves represent the case of $m_{\Delta}=1112$ MeV, blue ones $m_{\Delta}=1232$ MeV and the black ones $m_{\Delta}=1352$ MeV. The magenta box represent the constraint of ref.\cite{Podsiadlowski} on baryonic mass ($M_B=(1.366 - 1.375)M_{\odot}$) for Pulsar B of binary system PSR J0737-3039 with gravitational mass ($M_G=(1.249 \pm 0.001) M_{\odot}$) \cite{Burgay}.}
\protect\label{mgmb}
\end{figure}

 In fig. \ref{mgmb} shows the variation of gravitational mass with respect to baryonic mass for different parameterization of $\beta$ and $\chi$ with different $\Delta$ masses. As expected baryonic mass changes in a similar way as gravitational mass with the variation of $\beta$ and $\chi$. The inset of fig. \ref{mgmb} confirms that the constraint on baryonic mass from PSR J0737-3039 B ($M_B=(1.366 - 1.375)M_{\odot}$) \cite{Podsiadlowski} with corresponding maximum gravitational mass $M_G=(1.249 \pm 0.001) M_{\odot}$ \cite{Burgay} has been satisfied for all the variations of $\beta$ and $\chi$ for all the $\Delta$ masses. However, it is seen that for $m_{\Delta}=1352$ MeV (black curves) the constraint is just satisfied. For $m_{\Delta}=1232$ MeV (blue curves) and $m_{\Delta}=1112$ MeV (red curves) the constraint is gradually well satisfied with formation of more and more $\Delta$s.
 
 Overall it is seen that the effects of modified pressure are extremely important to achieve high mass NS configurations \cite{Ant} in the presence of exotic matter like the hyperons and $\Delta$s. It is noteworthy that these effects of modified pressure are particularly significant for massive NSs as for low mass NSs it is seen from figs. \ref{mr1112}, \ref{mr1232} and \ref{mr1352} that these effects gradually become less prominent and finally reduce to the normal GR solutions. With the application of the PTOV equations, there is an increase of maximum $\sim 20.5 \%$ in the gravitational mass of NSs for each $\\Delta$ mass. These modifications also help to satisfy the constraints imposed on NS properties like $R_{1.4}$ and $R_{1.6}$ from the detection of GW170817 for BNS merger \cite{Abbott,Fattoyev,Bauswein}. The constraint on baryonic mass of NSs obtained from PSR J0737-3039 \cite{Burgay,Podsiadlowski} is also satisfied.
 
 
\section{Summary and Conclusion}

The possibility of formation of $\Delta$ resonances in NSM in the presence of the hyperons in NSM is investigated in this work. As the $\Delta$s posses Breit-Wigner mass distribution with a specified width, I look for the effect of variation in $\Delta$ mass on the EoS and the resultant global properties of NSs. For this purpose the hyperon-meson couplings are calculated reproducing their potential depths in nuclear matter while the $\Delta$-meson couplings are chosen consistent to the prescriptions from QCD calculations \cite{qcd}. It is seen that the variation in $\Delta$ mass brings considerable changes to the formation and relative population of the different baryons and leptons. $\Delta$s, when taken to be lighter, their formation is greatly favored in NSM. In such case all the $\Delta$s are formed and they not only suppress the formation of all the hyperons but also become one of the major constituent of NSM along with the nucleons. The scenario, however, changes and only $\Delta^-$ and $\Delta^0$ are formed in NSM with much less concentration when the $\Delta$ mass is increased. The hyperons are then the predominant constituent in NSM along with the nucleons. It is seen that with further increase in $\Delta$ mass, they are totally restricted by the increased formation and concentration of the hyperons. The variation in formation of the $\Delta$s are seen to affect the NS properties greatly. With massive $\Delta$s, comparatively massive NS configurations are obtained. However, due to the formation of substantial amount of exotic matter like hyperons and the $\Delta$s, the EoS softens quite a lot. As a result the $\sim 2 M_{\odot}$ maximum mass constraint from PSR J0348+0432 \cite{Ant} is not satisfied with our model within the framework of GR even when the $\Delta$s are considered to be most massive and therefore the delta puzzle still remains unresolved in normal GR conditions. However, the estimates of $R_{1.4}$ and $R_{1.6}$ for all the $\Delta$ masses are consistent with the bounds on the same obtained from GW170817 data \cite{Abbott,Fattoyev,Bauswein}.

 To resolve the delta puzzle the effects of modified inertial pressure (via $\beta$) and self gravity (via $\chi$) (as suggested by \cite{Velten}) are incorporated in this work. This brings significant changes to the static properties of NS mostly the gravitational mass. It is seen that all the configurations of $\beta$ and $\chi$ used to modify normal GR conditions increases the mass of NS than that obtained with normal GR configuration ($\beta$ = 1 and $\chi$ = 1) thereby resolving the delta puzzle. The different configurations of $\beta$ and $\chi$ show that both inertial pressure and self-gravity must be very low in order to explain massive NS configurations. Even for each modification, the recent constraints from GW170817 on NS properties like the estimates of $R_{1.4}$ and $R_{1.6}$ are also satisfied. As the possible presence of exotic matter like the hyperons and the $\Delta$s within dense cores of massive NSs cannot be neglected, these effects of modified GR conditions in terms of pressure thus become extremely important and effective in explaining the structure and composition of such massive NSs like PSR J1614-2230 and PSR J0348+0432. It is clearly reflected that for these high mass NSs, the inertial pressure and the self-gravity may get modified which can be explained with these modified GR conditions even with soft EoS in presence of exotic matter like the hyperons and $\Delta$s. However, recent literature presents a huge variety of EoS from various theoretical models and thus a great deal of uncertainty is related to the EoS of NSM. Therefore, the extent to which the massive NSs can bring modifications to these effects of pressure still remains undetermined and a model dependent finding. However, within the framework of our model, the concerned EoS obtained with hyperons and $\Delta$s, it is seen that massive NSs may well constrain gravity along with EoS. The modified GR conditions may play important role in determining more accurate mass of NSs when a particular EoS is chosen. Apart from the maximum mass constraints from PSR J1614-2230 and PSR J0348+0432, the theory of parameterized hydrostatic equilibrium successfully satisfies the constraints on $R_{1.4}$ and $R_{1.6}$ from GW170817 and also that on baryonic mass of NSs obtained from PSR J0737-3039.

\section*{Acknowledgments}
 
I am thankful to Dr. Kinjal Banerjee and Dr. T. K. Jha for their useful suggestions.


\end{document}